\documentclass[reprint,twocolumn,aps,amsmath,amssymb,prb,superscriptaddress,floatfix]{revtex4}

\usepackage{graphicx}
\usepackage{longtable}
\usepackage{amsmath}\usepackage{bm}

\begin{document}

\title{Dynamics of a localized spin excitation close to the spin-helix regime}

\author{G.~Salis}
\email{gsa@zurich.ibm.com}
\affiliation{IBM Research--Zurich, S\"aumerstrasse 4, 8803 R\"uschlikon, Switzerland}

\author{M.~P.~Walser}
\affiliation{IBM Research--Zurich, S\"aumerstrasse 4, 8803 R\"uschlikon, Switzerland}

\author{P.~Altmann}
\affiliation{IBM Research--Zurich, S\"aumerstrasse 4, 8803 R\"uschlikon, Switzerland}

\author{C.~Reichl}
\affiliation{Solid State Physics Laboratory, ETH Zurich, 8093 Zurich, Switzerland}

\author{W.~Wegscheider}
\affiliation{Solid State Physics Laboratory, ETH Zurich, 8093 Zurich, Switzerland}

\begin{abstract}
The time evolution of a local spin excitation in a (001)-confined two-dimensional electron gas subjected to Rashba and Dresselhaus spin-orbit interactions of similar strength is investigated theoretically and compared with experimental data. Specifically, the consequences of the finite spatial extension of the initial spin polarization is studied for non-balanced Rashba and Dresselhaus terms and for finite cubic Dresselhaus spin-orbit interaction. We show that the initial out-of-plane spin polarization evolves into a helical spin pattern with a wave number that gradually approaches the value $q_0$ of the persistent spin helix mode. In addition to an exponential decay of the spin polarization that is proportional to both the spin-orbit imbalance and the cubic Dresselhaus term, the finite width $w$ of the spin excitation reduces the spin polarization by a factor that approaches $\exp(-q_0^2 w^2/2)$ at longer times.
\end{abstract}

\maketitle

\section{Introduction}

The spin-orbit interaction (SOI) in atoms is a relativistic correction of the orbital energy levels. It can be understood as the interaction of the electron spin with a magnetic field that has its origin in the Lorentz-transformed electrostatic field of the atomic core. In a crystalline solid, this interaction influences the band energies; specifically, it leads to a spin splitting of electronic states with finite momentum in crystals with an inversion asymmetry. A crystallographic inversion asymmetry as it exists, e.g., in bulk zincblende semiconductors leads to the Dresselhaus SOI~\cite{Dresselhaus1955} for conduction-band electrons. In addition, a Rashba SOI can be induced~\cite{Vasko1979, Bychkov1984} in a layered structure by applying an electric field perpendicular to the layers. For electrons confined in a quantum well, the Dresselhaus spin splitting depends on the quantum confinement~\cite{Dyakonov1986}. For a typical material like a GaAs-based quantum well, it is  on the order of 100\,$\mu$eV\cite{Walser2012b} at the Fermi energy, which translates into a large effective magnetic field of 5-10 T. This makes SOI an interesting tool for coherent manipulation of electronic spin states in solids. On the other hand, the large field is a significant source for the decay of the spin polarization by the so-called Dyakonov-Perel mechanism~\cite{Dyakonov1972}. It has been shown theoretically~\cite{Schliemann2003,Bernevig2006} that by balancing the Dresselhaus and the Rashba contribution to SOI, the interaction attains a special symmetry with respect to the size and direction of the electron momentum and leads to the preservation of a helical spin mode. For (001)-oriented GaAs-based quantum wells, the spin polarization of this helical mode rotates about an in-plane axis when the position is varied along the perpendicular in-plane direction. The measured decay rate of imprinted spin gratings of variable wave number~\cite{Koralek2009} follows the theoretical prediction with a minimum decay rate at the wave number $q_0$ of a persistent spin helix~\cite{Bernevig2006}.


In a typical experimental configuration, spin polarization is injected locally into the non-magnetic semiconductor by, e.g., spin injection contacts~\cite{Lou2007} or optical orientation~\cite{Wunderlich2010}. It is interesting to consider the evolution of such spin polarization into a spin helix pattern. In general, the initial spin polarization can be described by a superposition of helical spin modes with different wave vectors $\mathbf q = (q_x,q_y)$ and decay rates $\Gamma(q_x,q_y)$~\cite{Yang2010}. Measurements show that for a spatially confined spin excitation, this superposition evolves into a helical spin pattern that diffusively expands with time and whose wave number evolves towards $q_0$~\cite{Walser2012}. For a balanced SOI, where the Rashba SOI coefficient $\alpha$ is equal to $\beta=\beta_1-\beta_3$ (with $\beta_1$ and $\beta_3$ the linear and cubic Dresselhaus coefficients as defined below), the evolution of a spatially delta-shaped spin excitation can be analytically described by the product of a Gaussian function and a helical mode with wave number $q_0$~\cite{Yang2010}. In real situations, the spin excitation has a finite extension and SOI may not be balanced. To understand the experimentally observed spin dynamics, it is important to analyze to what extend such non-idealities alter this description.

Here, we theoretically study the effects that occur under realistic experimental conditions where (i) the Rashba and Dresselhaus SOI are not balanced and cubic Dresselhaus terms are present, (ii) the initial out-of-plane spin polarization has a finite spatial extension along the direction $y$ of the helical spin precession but is constant along the perpendicular direction $x$, and (iii) spin polarization is localized in both in-plane directions. The model derived bridges the gap between a delta-shaped and a spatially broad excitation, i.e., between the formation of a long-lived helical spin mode and a spatially homogeneous spin decay described by the Dyakonov-Perel mechanism. We find an exponential decay with a rate proportional to $D_s ((\alpha-\beta)^2+3\beta_3^2)$, where $D_s$ is the spin diffusion constant. An initial spin polarization $S_z(y)$ with finite extension along $y$ will be further reduced because of the transition of the relevant wave numbers from zero to $q_0$. This latter effect is also responsible for a gradual decrease of the period of the evolving helical spin polarization. Analytical results are derived on the assumption of an initial spin polarization $S_z(y)$ that is independent of $x$. If the initial spin polarization is also confined along $x$, the spin eigenmodes and their dispersion exhibit anticrossings for $\alpha\neq\beta$. We show that the localization along $x$, however, does not appreciably alter the behavior of the spin evolution apart from a trivial diffusive expansion along that direction. Finally, we verify the predicted transients of the polarization amplitude and the mode wave number in an experiment in which spin polarization in a GaAs quantum well is initialized in confined areas of different extensions.

\section{Model}
\subsection{Spin-Diffusion equation}

We first define the coordinate system and the spin-orbit coefficients. We consider a two-dimensional electron gas confined along the [001] crystalline direction $z$ of a zincblende crystal, such as GaAs. We define the two in-plane directions $x$ and $y$ along $[1\overline{1}0]$ and $[110]$, respectively. The Hamiltonian for an electron with in-plane wave vector $\mathbf k = (k_x,k_y)$ and effective mass $m^*$ is

\begin{equation}
H=\frac{\hbar^2 k^2}{2m^*}+\frac{\hbar}{2}\mathbf{\Omega}_{\textrm SO}\cdot\boldsymbol{\sigma},
\end{equation}

with the SOI defined by

\begin{equation}
\mathbf{\Omega}_\textrm{SO} = \frac{2}{\hbar}
\begin{pmatrix}
\left(\alpha+\beta_1+2\beta_3\frac{k_x^2-k_y^2}{k^2}\right)k_y\\
\left(-\alpha+\beta_1-2\beta_3\frac{k_x^2-k_y^2}{k^2}\right)k_x\\
0
\end{pmatrix}.
\label{eq:Hamiltonian}
\end{equation}

The spin is represented by the three Pauli matrices $\boldsymbol\sigma=(\sigma_1, \sigma_2, \sigma_3)$.
The Dresselhaus SOI coefficients $\beta_1$ and $\beta_3$ are related to the bulk Dresselhaus coefficient $\gamma$ by $\beta_1=-\gamma \langle k_z^2 \rangle$ and $\beta_3=-\gamma k^2/4$. The expectation value of $k_z^2$ with respect to the QW ground-state envelope wave-function is denoted by $\langle k_z^2 \rangle$. Whereas $\alpha$ and $\beta_1$ give rise to SOI that is linear in $k=|\mathbf k|$,  the cubic coefficient $\beta_3$ itself depends quadratically on $k$ and in total accounts for SOI that is cubic in $k$. We consider a degenerate electron gas with $k_\textrm{B}T\ll \hbar^2 k_\textrm{F}^2/(2m^*)$, where $k_\textrm{F}$ is the Fermi wave number and $k_\textrm{B}$ the Boltzmann constant. Also, we assume that the SOI is small compared with the Fermi energy and that the initial spin polarization density is small compared with the electron sheet density. The relevant electronic states are then centered at the Fermi energy, and $k$ can be replaced by $k_{\textrm F}$ in Eq.~(\ref{eq:Hamiltonian}) and in the definition of $\beta_3$.

We investigate a situation close to the balanced SOI of a perfect spin helix and with the same signs for $\alpha$ and $\beta$. We characterize this condition by $r^2\ll 1$, introducing the parameters $r_1=(\alpha-\beta)/(\alpha+\beta)$, $r_2=\beta_3/(\alpha+\beta)$, and $r^2=r_1^2+r_2^2$. To describe the evolution of a general spin polarization $\boldsymbol \rho(x,y)$ in direct space, it is favorable to consider its Fourier components $\widehat{\boldsymbol \rho}(q_x, q_y)$ that harmonically oscillate in space and time:

\begin{equation}
\label{eq:timeevol1}
\boldsymbol \rho(x,y,t)=a \widehat{\boldsymbol \rho}(q_x,q_y)e^{iq_x x + iq_y y - i\omega t}.
\end{equation}

For the Fourier-space spin polarization $\widehat{\boldsymbol{\rho}}$, it is possible to derive a spin-diffusion equation in the presence of SOI. Using a density-matrix response function with standard perturbation theory~\cite{Burkov2004,Stanescu2007,Liu2012}, or starting from a semiclassical spin kinetic equation~\cite{Lueffe2011}, the following equation is obtained:

\begin{equation}\label{eq:Diffeq}
\left(-i\omega + D_s(q_x^2+q_y^2+q_0^2 \tilde{\boldsymbol D})\right)
\widehat{\boldsymbol \rho}=0.
\end{equation}

We defined the spin helix wave number $q_0=\frac{2m^*}{\hbar^2}(\alpha+\beta)$ and the spin diffusion constant $D_s=\hbar^2 k_{\textrm F}^2 \tau/(2 {m^*}^2)$ that depends on the effective electron momentum scattering time $\tau$. Note that $\tau$ contains also contributions from electron-electron scattering~\cite{Weber2005}, which is not the case for charge diffusion. The term $q_0^2 D_s \tilde{\boldsymbol{D}}$ accounts for the spin dynamics related to SOI. The diagonal elements of this matrix yield the Dyakonov-Perel dephasing rates. The non-diagonal terms arise from correlations between the momentum and the spin-orbit field and drive the helical spin modes. Equation~(\ref{eq:Diffeq}) is valid in the weak spin-orbit regime where the scattering time is small compared to the spin precession period, $\Omega_{\textrm{SO}}\tau \ll 1$. It does not account for additional spin scattering mechanisms as Elliott-Yafet~\cite{Liu2012} or Bir-Aronov-Pikus~\cite{Volkl2011}. The matrix $\tilde{\boldsymbol{D}}$ is given by

\begin{equation}\label{eq:matrixD}
\tilde{\boldsymbol{D}}=
\begin{pmatrix}
r_1^2+r_2^2 && 0 && -ir_1\frac{2 q_x}{q_0}\\
0 && 1+r_2^2 && -i\frac{2 q_y}{q_0}\\
ir_1\frac{2 q_x}{q_0} && i\frac{2q_y}{q_0} && 1+r_1^2+2r_2^2
\end{pmatrix}.
\end{equation}

By determining the eigenvectors $\widehat{\boldsymbol\rho}_n(q_x,q_y)$ and eigenvalues $\lambda_n$ of $\tilde{\boldsymbol{D}}$, one obtains for each pair $(q_x, q_y)$ three eigenmodes $n=1,2,3$  that solve Eq.~(\ref{eq:Diffeq}) and that decay exponentially with a rate

\begin{equation}
\label{eq:dispersiongeneral}
i\omega_n=D_s(q_x^2+q_y^2+q_0^2 \lambda_n).
\end{equation}

The time evolution of an arbitrary spin polarization $\boldsymbol\rho(x,y,t)$ in direct space can then be expressed in terms of these eigenmodes by using the Fourier integral

\begin{equation}
\label{eq:timeevol2}
\boldsymbol\rho(x,y,t)=\int_{-\infty}^{\infty}\int_{-\infty}^{\infty}{\sum_{n=1}^{3} a_n \widehat{\boldsymbol\rho}_n e^{-i\omega_n t+i q_x x+i q_y y}dq_x dq_y}.
\end{equation}

Here, $a_n(q_x, q_y)$ are the amplitudes of the excited eigenmodes.

\subsection{Evolution of a spin excitation with finite extension along the helix direction}

We first discuss the situation where only eigenmodes with $q_x=0$ are excited, i.e. where the initial spin polarization does not vary as a function of $x$. Setting $q_x=0$ in Eq.~(\ref{eq:matrixD}), the eigenvalues of $\tilde{\boldsymbol{D}}$ are~\cite{Liu2012}

\begin{equation}
\lambda_1=r_1^2+r_2^2=r^2
\end {equation}

and

\begin{equation}
\label{eq:eigenvalues}
\lambda_{2,3}=1+\frac{1}{2}r_1^2+\frac{3}{2}r_2^2 \pm
\frac{1}{2}\sqrt{16\frac{q_y^2}{q_0^2}+r^4}.
\end{equation}

$\lambda_1$ corresponds to a mode with a unidirectional spin polarization along the $x$-direction. $\lambda_2$ and $\lambda_3$ are the eigenvalues of two helical modes with opposite helicity.

In the special situation of a perfect spin helix ($r_1=0$, $r_2=0$), one obtains~\cite{Yang2010} $\lambda_{2,3}=1\pm2q_y/q_0$, and thus $i\omega_{2,3}=D_s\left(q_y \pm q_0\right)^2$. The spin decay of mode 3 is completely suppressed for $q_y=q_0$. Note that the same is true for mode 2 at $q_y=-q_0$. As we will see below, also the eigenvectors of modes 2 and 3 interchange when the sign of $q_y$ is inverted.

In the general case where $r_1 \neq 0$ or $r_2 \neq 0$,  the square root term in Eq.~(\ref{eq:eigenvalues}) can be approximated by $2q_y/q_0$ as long as $q_y\gg r^2 q_0/4$. This yields

 \begin{equation}
 \label{eq:eigenvalueapprox}
 \lambda_{2,3}=1+\frac{1}{2}r_1^2+\frac{3}{2}r_2^2 \pm 2q_y/q_0.
 \end{equation}

In the other limit where $q_y=0$, $\lambda_2$ and $\lambda_3$ are split by $r^2$, which is however much smaller than $\lambda_2 \approx \lambda_3 \approx 1$. As a consequence, Eq.~(\ref{eq:eigenvalueapprox}) is a good approximation for all $q_y$. The dispersion of modes 2 and 3 can thus be written as

\begin{equation}
\label{eq:dispersion}
i\omega_{2,3}=D_s\left(q_y \pm q_0\right)^2+\Gamma_s,
\end{equation}

with

\begin{equation}
\label{eq:Gammas}
\Gamma_s=\frac{1}{2}D_s q_0^2\left(r_1^2+3r_2^2\right).
\end{equation}

For $r_1\neq 0$ or $r_2\neq0$, in addition to the exponential decay rate $D_s (q_y\pm q_0)^2$, a decay with rate $\Gamma_s$ occurs. Note that in $\Gamma_s$, the cubic Dresselhaus term $r_2^2$ is weighted more (by a factor of three) than the imbalance term $r_1^2\propto (\alpha-\beta)^2 $. The proportionality of $\Gamma_s$ to $3\beta_3^2$ for $\alpha=\beta$ was derived in Ref.~\onlinecite{Yang2010}.

\begin{figure}[ht]
\includegraphics[width=85mm]{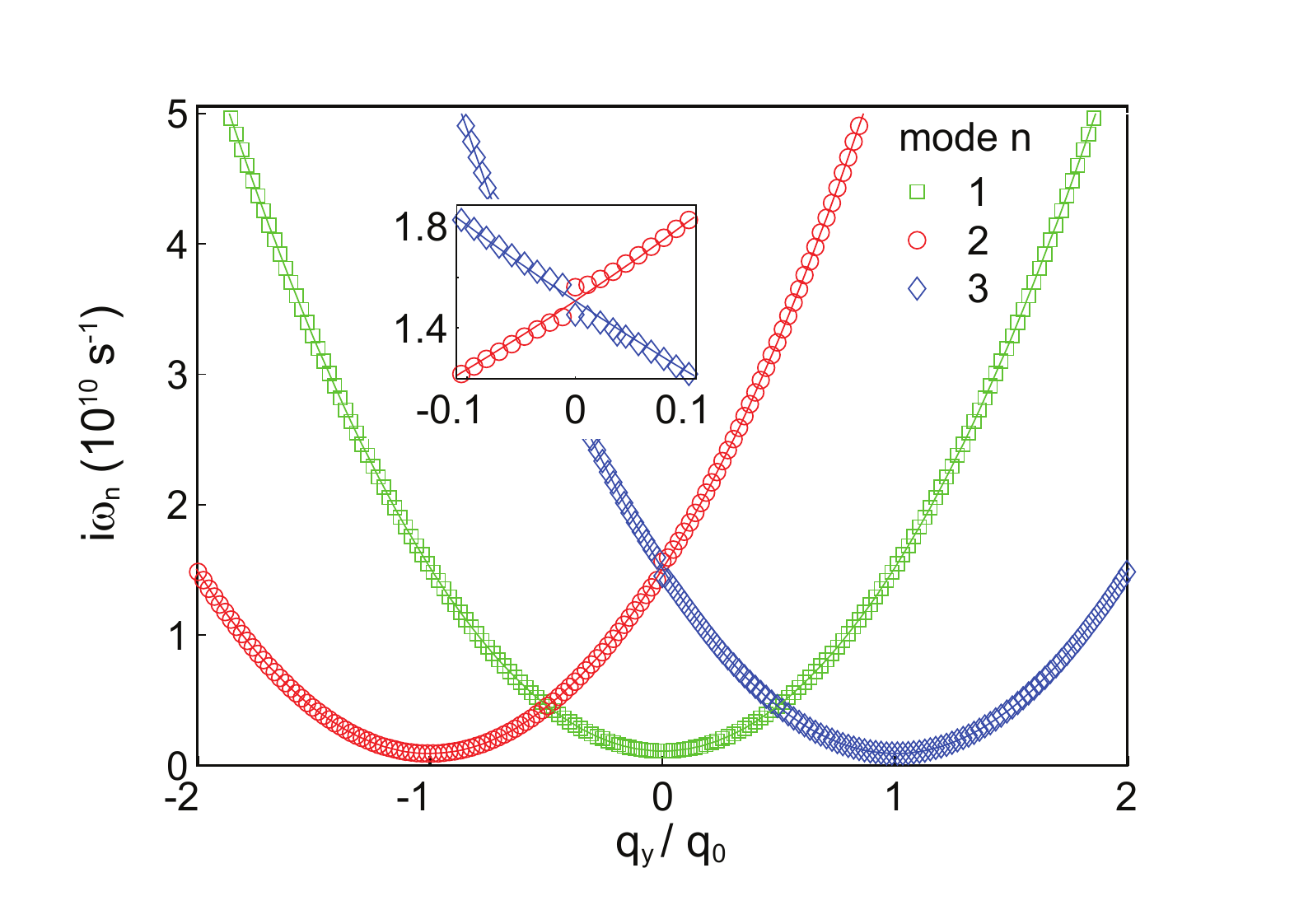}
\caption{\label{fig:fig1} Dispersion of the eigenmodes of the spin diffusion equation for $q_x=0$ with parameters as given in the main text. The quantity $i\omega_n$ indicates the spin decay rate of mode $n$ and is shown as a function of the wave number $q_y$. At $q_y=q_0$ ($q_y=-q_0$), the decay rate of mode 3 (2) is at its minimum, corresponding to the persistent spin helix. Symbols correspond to the exact values calculated from Eqs.~(\ref{eq:dispersiongeneral}) and (\ref{eq:eigenvalues}), whereas the solid lines are the approximations based on Eqs.~(\ref{eq:dispersion}) and (\ref{eq:Gammas}). For $r^2\ll1$, the two solutions only deviate around $q_y=0$ where a small splitting of modes 2 and 3 is neglected in the approximation (see inset). }
\end{figure}

For illustration purpose, we will assume the following parameters for the SOI: $\alpha=1.7\cdot10^{-13}$\,eVm, $\beta_1=3.4\cdot10^{-13}$\,eVm, $\beta_3=0.7\cdot10^{-13}$\,eVm, $\tau=0.5$\,ps and an electron sheet density of $n_s=5\cdot10^{15}$\,m$^{-2}$. Figure~\ref{fig:fig1} compares the mode dispersion that follows from Eqs.~(\ref{eq:dispersiongeneral}) and (\ref{eq:eigenvalues}) with the approximations Eqs.~(\ref{eq:dispersion}) and (\ref{eq:Gammas}). Even for the relatively large deviation from balanced SOI ($r^2=0.077$), the splitting of modes 2 and 3 at $q_y=0$ is small (inset of Fig.~\ref{fig:fig1}) and the dispersion is well approximated by the parabolic functions given in Eq.~(\ref{eq:dispersion}).

We consider a spin polarization that at time $t=0$ is oriented along the $z$-direction. The spatial distribution is assumed to be uniform along $x$ and Gaussian along $y$ with a width of $w_y$, and an amplitude $A$: $\boldsymbol\rho(x,y,0)=(0,0,A \exp(-y^2/2w_y^2))$. This corresponds to a spin polarization in Fourier space at $t=0$ of

\begin{equation}
a \widehat{\boldsymbol\rho}(q_y)=
\begin{pmatrix}
0\\
0\\
A\frac{w_y}{\sqrt{2\pi}}\exp\left(-\frac{q_y^2 w_y^2}{2}\right)
\end{pmatrix}.
\end{equation}

Because of the uniform distribution along $x$, we can omit the Fourier transformation along $q_x$ in Eq.~(\ref{eq:timeevol2}). For each $q_y$, $a \widehat{\boldsymbol \rho}$ is decomposed into three eigenvectors $\widehat{\boldsymbol\rho}_n$ with amplitudes $a_n$. With the initial spin polarization along $z$, only modes 2 and 3 are excited; the polarization of mode 1 is uniformly pointing along $x$. Calculating the eigenvectors of $\tilde{\boldsymbol{D}}$ and using the approximation of Eq.~(\ref{eq:dispersion}), the two modes are given by

\begin{widetext}
\begin{equation}
a_{2,3}\widehat{\boldsymbol\rho}_{2,3}=
\frac{Aw_y}{2\sqrt{2\pi}}\exp\left(-\frac{q_y^2 w_y^2}{2}\right)\left(1\pm r^2\frac{q_0}{4 q_y}\right)
\begin{pmatrix}
0\\
\pm i(1 \mp r^2 \frac{q_0}{4 q_y})\\
1
\end{pmatrix}.
\end{equation}
\end{widetext}

\begin{figure}[ht]
\includegraphics[width=85mm]{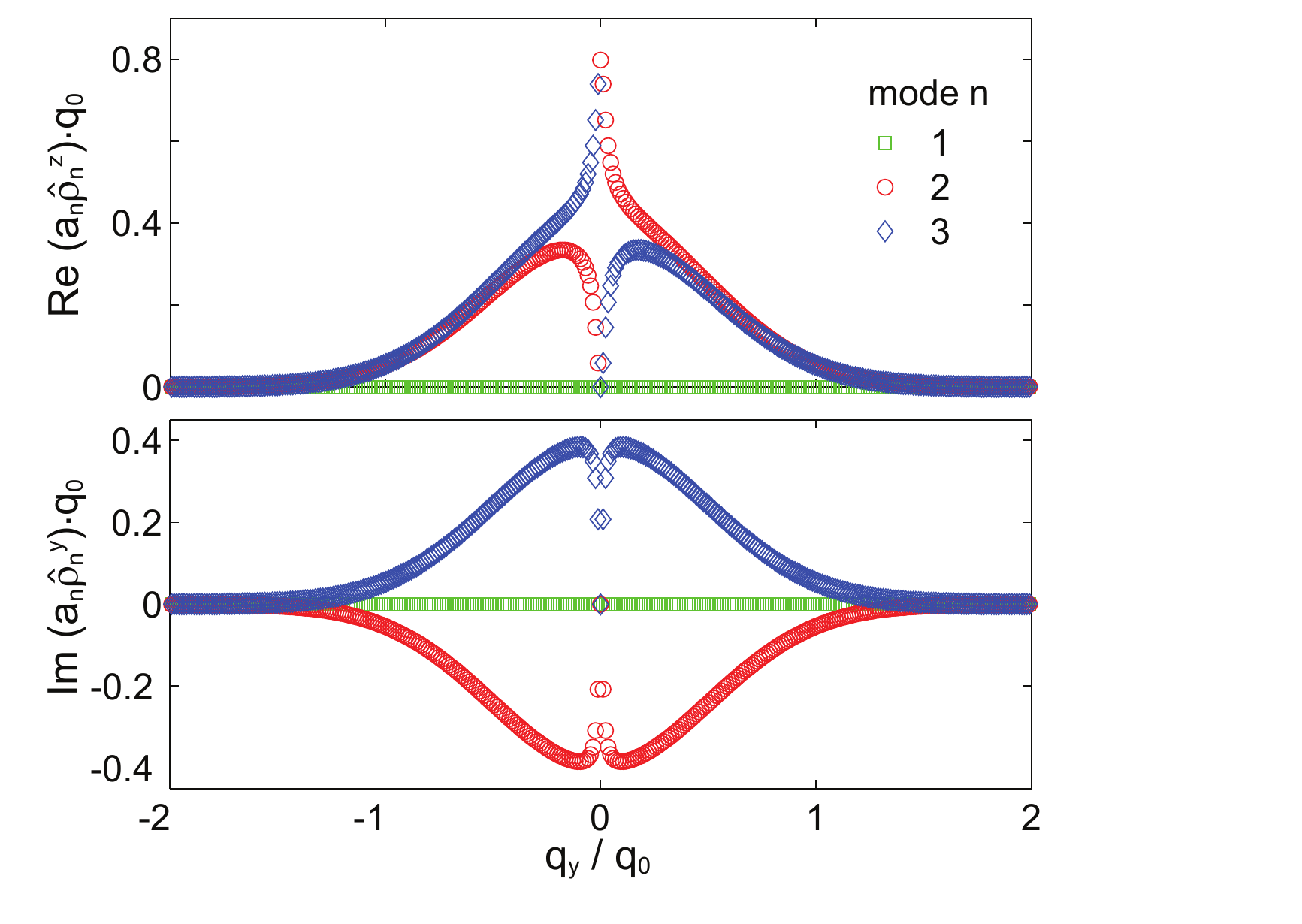}
\caption{\label{fig:fig2} Spin polarization of the three eigenmodes in $q$-space directly after excitation by a spin polarization oriented along $z$ and with a Gaussian profile along the $y$-direction. The initial width of the profile is set to $w_y=2/q_0$. Only the real part of the $z$-components (a) and the imaginary part of the $y$-components (b) of modes 2 and 3 have a finite size. Mode 1 is not excited.}
\end{figure}

Figure~\ref{fig:fig2} displays the real and imaginary parts of $a_n\widehat{\boldsymbol\rho}_n$ versus $q_y$. The $z$-component of mode 3 has a positive real value and the $y$-component a positive imaginary one. In direct space, this corresponds to the $z$-component being proportional to $\cos q_y y$, and the $y$-component to $-\sin q_y y$, see Eq.~(\ref{eq:timeevol1}). This constitutes a helical spin mode where the spin polarization rotates counterclockwise in the $y-z$ plane when moving towards the positive $y$ axis (for a positive $q_y$ and observed from the positive $x$-axis). In contrast, the negative imaginary value of $a_2\widehat{\rho}_2^y$ leads to a clockwise rotating helical spin mode. Note that if the sign of $q_y$ is reversed, also the helicity of the respective mode is flipped.

\begin{figure}[ht]
\includegraphics[width=75mm]{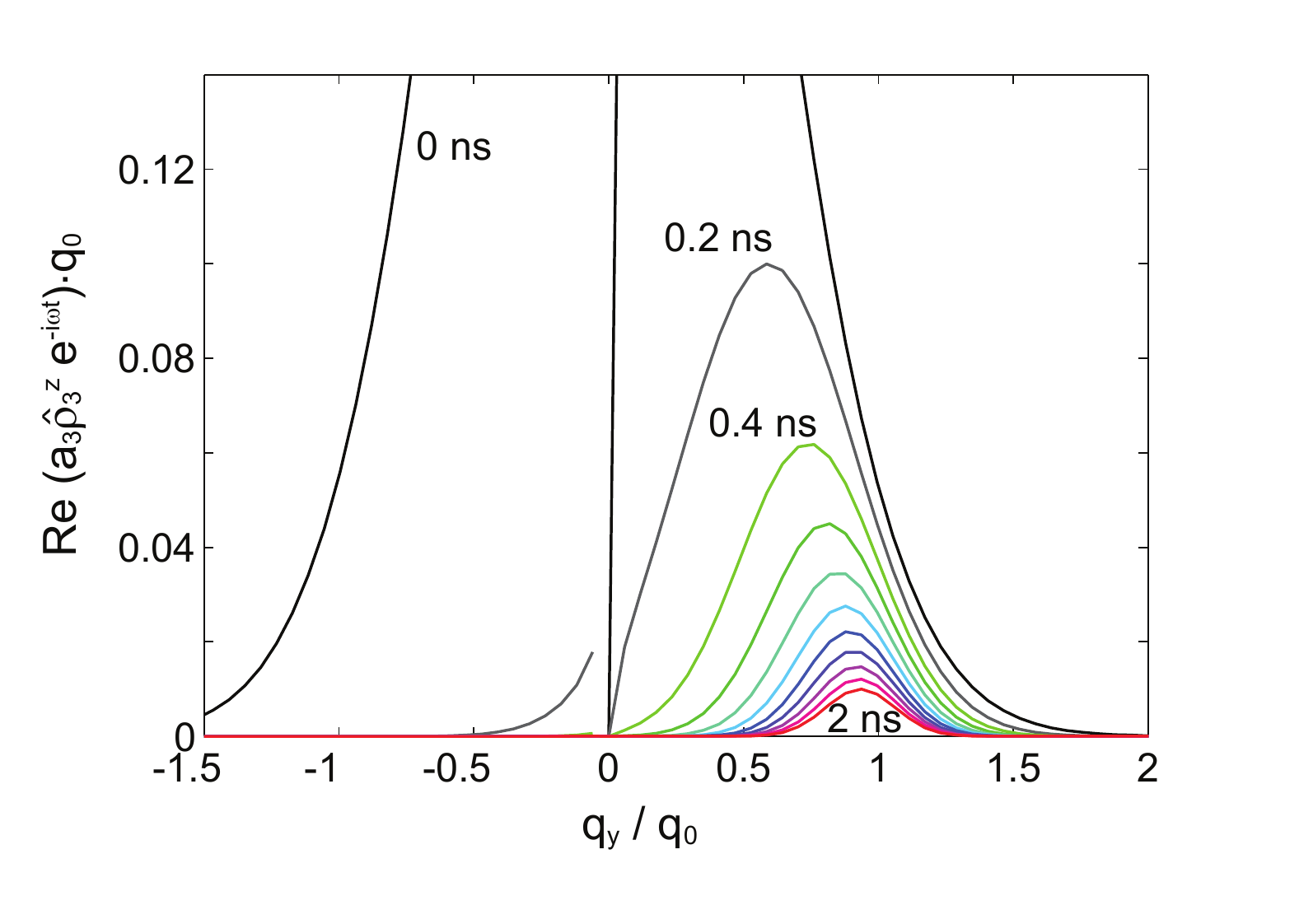}
\caption{\label{fig:fig3} Time evolution of mode 3 in $q$-space after the same excitation as shown in Fig.~\ref{fig:fig2}. With time, the excitation centered around $q_y=0$ shifts its weight towards $q_y=q_0$, where the decay rate is smallest, transforming the initial spin polarization along $z$ into a helical spin mode.}
\end{figure}

Each mode $\widehat{\boldsymbol \rho}_n(q_x,q_y)$ decays exponentially with a decay rate $i\omega_n(q_x,q_y)$. In Fig.~\ref{fig:fig3} we plot the time evolution of mode 3 exemplarily. As mode 3 has a minimum decay rate at $q_y=q_0$, the original excitation centered at $q_y=0$ will shift with time towards $q_y=q_0$. In the same way, the weight of mode 2 will displace towards $-q_0$.

The spin dynamics in direct space is found from Eq.~(\ref{eq:timeevol2}), where we consider only the physically meaningful real part of $\rho$. Making use of the symmetry of the mode dispersion with respect to $q_y=0$, we obtain
\begin{widetext}
\begin{equation}
\label{eq:mode}
\boldsymbol\rho=\frac{Aw_y}{\sqrt{2\pi}}\int_{-\infty}^{\infty}{e^{-q_y^2 w_y^2/2-\Gamma_s t- D_s(q_0-q_y)^2 t}
\begin{pmatrix}
0\\
\sin q_y y\left(1-r^4\frac{q_0^2}{16q_y^2}\right)\\
\cos q_y y\left(1-r^2\frac{q_0}{4q_y}\right)
\end{pmatrix}
}dq_y.
\end{equation}
\end{widetext}

The terms of the integrand that contain $r$ are only relevant around $q_y=0$ and can be neglected after integration because they are odd in $q_y$. This results in the following expression for the spin polarization in direct space:

\begin{widetext}
\begin{equation}
\label{eq:modeapprox}
\boldsymbol\rho(x,y,t)=A \frac{w_y}{w_y'}\exp\left({-\frac{y^2}{2{w_y'}^2}-D_s q_0^2\frac{w_y^2}{w_y'^2}t-\Gamma_s t}\right)
\begin{pmatrix}
0\\
\sin q_0' y \\
\cos q_0' y
\end{pmatrix}.
\end{equation}
\end{widetext}

The spin polarization establishes a helical oscillation with wave number $q_0'=\left(1-\frac{w_y^2}{w_y'^2}\right)q_0$. The envelope of this oscillation is given by a Gaussian distribution of width $w_y'=\sqrt{w_y^2+2 D_s t}$. Equation~(\ref{eq:modeapprox}) therefore describes the following modifications compared with a delta-shaped excitation: (1) The amplitude of the helical state decays not only with rate $\Gamma_s$, but also with an additional time-varying rate $D_s q_0^2w_y^2/w_y'^2$. (2) The spatial oscillation period of the spin polarization decreases with time and approaches $2\pi/q_0$ only asymptotically. (3) The diffusive expansion reduces the signal proportional to $w_y/w_y'=1/\sqrt{1+2D_s t/w_y^2}$, instead of just proportional to $1/\sqrt{t}$.

For $t\gg w_y^2/2D_s$, $q_0'\approx q_0$ and the time-varying decay rate suppresses the spin polarization by a constant factor of $\exp({-q_0^2 w_y^2/2})$. This means that if $w_y\ll 1/q_0$, then the evolution of $\boldsymbol \rho$ is equivalent to that of a delta-shaped excitation, at least for times $t>(D_s q_0^2)^{-1}$. For larger widths or shorter times, the modifications discussed above need to be considered to interpret experimental data.

\subsection{Spin excitation with finite extension along two directions}

If the initial spin polarization is localized along both the $x$- and the $y$-direction, $q_x$ cannot be set to zero in $\tilde{\boldsymbol D}$, and the eigenvalues differ from Eq.~(\ref{eq:eigenvalues}). However, the matrix elements that contain $q_x$ are much smaller than those with $q_y$ because $r_1\ll 1$. It is therefore tempting to use the same mode spectrum as for $q_x=0$, but to include the nonzero $q_x$ in Eq.~(\ref{eq:dispersiongeneral}). For a Gaussian excitation of width $w_x=w_y=w$, this yields the following result

\begin{widetext}
\begin{equation}
\label{eq:2Dmodeapprox}
\boldsymbol\rho(x,y,t)=A \frac{w^2}{w'^2}\exp\left({-\frac{x^2+y^2}{2{w'}^2}-D_s q_0^2\frac{w^2}{w'^2}t-\Gamma_s t}\right)
\begin{pmatrix}
0\\
\sin q_0' y \\
\cos q_0' y
\end{pmatrix},
\end{equation}
\end{widetext}

with

\begin{equation}
\label{eq:wtime}
w'^2=w^2+2 D_s t
\end{equation}

and

\begin{equation}
\label{eq:q0time}
q_0'=\left(1-\frac{w^2}{w'^2}\right)q_0.
\end{equation}

\begin{figure}[ht]
\includegraphics[width=85mm]{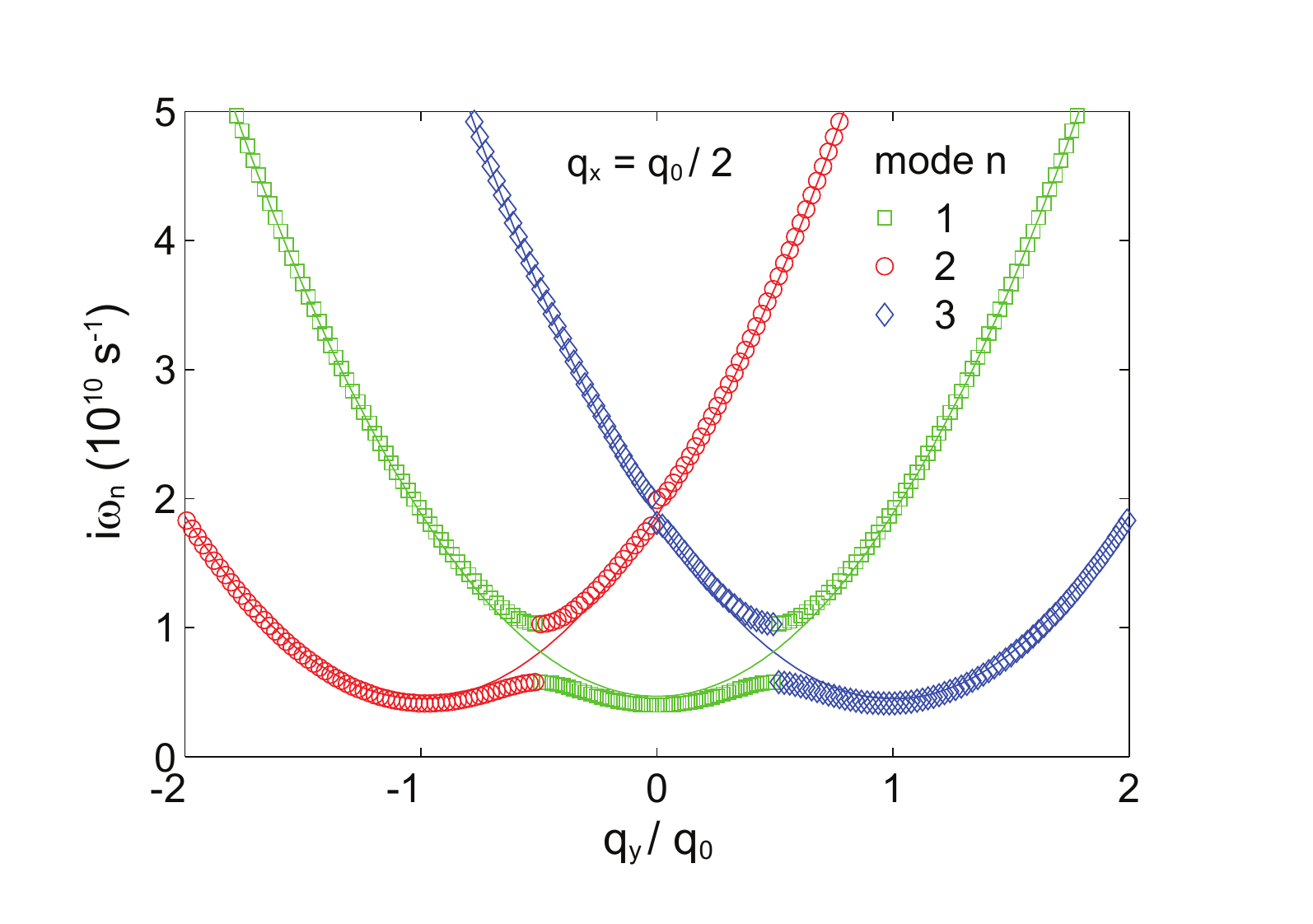}
\caption{\label{fig:fig5} Dispersion of the eigenmodes at $q_x=0.5 q_0$. Independent of $q_y$, the spin decay rate is increased by $D_s q_x^2$. In addition, the finite $q_x$ component leads to an anticrossing of the modes around $q_y=\pm0.5q_0$.}
\end{figure}

\begin{figure}[ht]
\includegraphics[width=85mm]{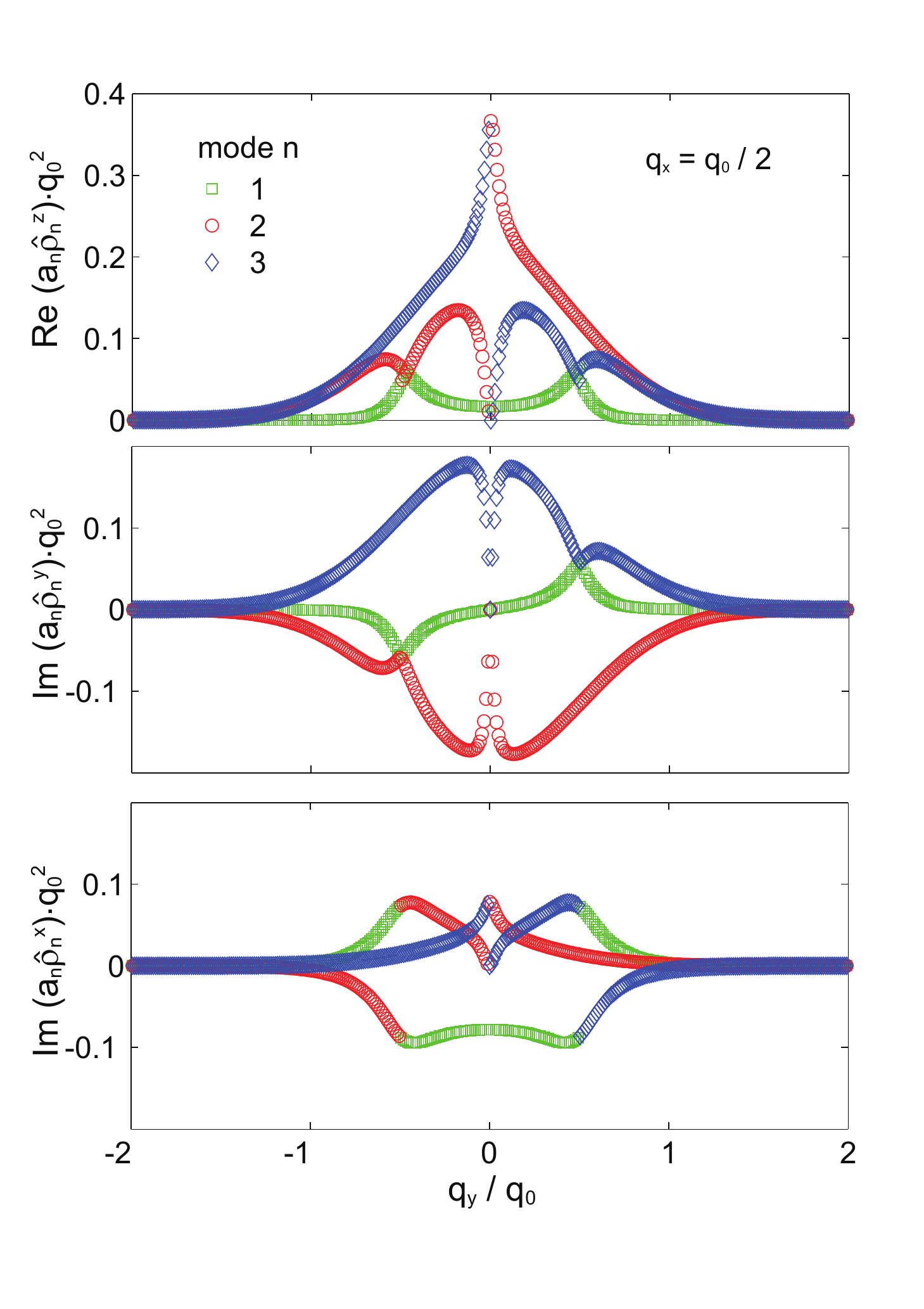}
\caption{\label{fig:fig6} Spin polarization of the three eigenmodes excited by a spin polarization along $z$ with a Gaussian profile of width $w=2/q_0$ along both the $x$- and $y$-directions. Modes are shown as a function of $q_y/q_0$ at $q_x=q_0/2$. In contrast to the case where the spin polarization is independent of the position $x$, all modes have finite spin components along all three directions, and all modes are excited simultaneously. The real parts of the $x$ and $y$ components, as well as the imaginary part of the $z$ component are not excited.}
\end{figure}

In the exact solution that takes nonzero $q_x$ in $\tilde{\boldsymbol D}$ into account, anticrossings of the eigenvalues occur close to $q_y=q_0/2$. Figure~\ref{fig:fig5} shows numerically calculated (exact) values of the mode dispersion $\omega_n$ for $q_x=q_0/2$. Independent of $q_y$, diffusion along the $x$ direction increases the decay rate by $D_s q_x^2$, weakening the relevance of such avoided crossings for the overall spin dynamics. In Fig.~\ref{fig:fig6}, the calculated nonzero components of $\widehat{\boldsymbol \rho}_n$ are shown for the specific case of $q_x=q_0/2$. All three eigenmodes are excited with finite components along all three directions. For $q_y>0$, the dispersion of mode 2 exhibits no anticrossing, and therefore the mode spectrum is excited similarly as for $q_x=0$, see Fig.~\ref{fig:fig2}. However, modes 1 and 3 anticross around $q_y=q_0/2$, and therefore those modes share the weight of the initial excitation. For $q_y<0$, mode 3 stays unaffected by the anticrossing, whereas modes 2 and 3 share the weight. Interestingly, also the $x$-components of all modes are excited. As required, the $x$-component of the sum of all modes is zero at time $t=0$, whereas a finite spin polarization along $x$ emerges at finite times because of the different decay times of the three modes.

\begin{figure}[ht]
\includegraphics[width=85mm,angle=270]{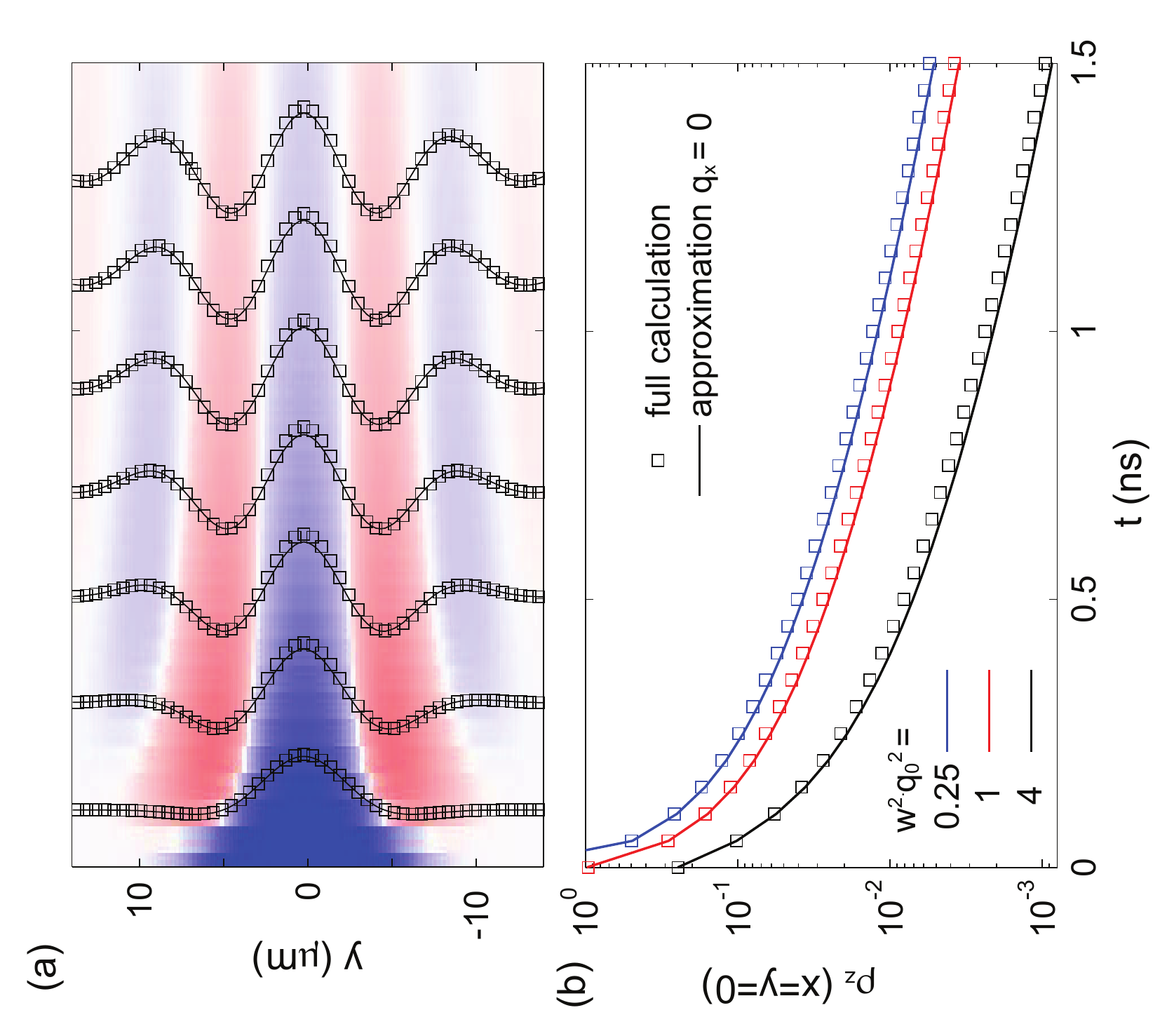}
\caption{\label{fig:fig4} Evolution of the spin polarization $\rho_z$ in direct space, after inital excitation of a Gaussian shape along $x$ and $y$. In (a), cross sections of $\rho_z(x=0,y)$ are shown at different times $t$ after excitation, with normalized amplitudes. Symbols correspond to the approximation where $q_x$ is set to zero in $\tilde{\boldsymbol D}$, whereas for the solid lines the full matrix was considered. In (b), the amplitude of $\rho_z$ at $x=y=0$ is shown versus time for the two cases and for different inital widths $w$.}
\end{figure}

Figure~\ref{fig:fig4} compares the difference in the evolution of spin polarization in direct space for the two cases, namely, whether the approximation $q_x=0$ is used in the diffusion matrix $\tilde{\boldsymbol D}$ or not. In the former case, $\rho_z$ is calculated using Eq.~(\ref{eq:2Dmodeapprox}), in the latter case, $\rho_z$ is numerically obtained by a Fourier transformation of the eigenmodes according to Eq.~(\ref{eq:timeevol2}). In Fig.~\ref{fig:fig4}(a), the evolution of the local excitation into the helical spin pattern is visible in the color-scale representation of the spin polarization $\rho_z(x=0,y,t)$. Cross sections of $\rho_z(y)$ at different times $t$ are superposed as data points (full calculation) and as solid lines (approximation of $q_x=0$). No significant difference is seen. In Fig.~\ref{fig:fig4}(b), the amplitude of $\rho_z(0,0,t)$ is shown versus $t$ on a logarithmic plot for three different widths of the initial spin polarization. Supposing that an equal number of spin-polarized electrons are excited in the three cases, the amplitude $A$ is scaled with $1/w^2$ in the plot. The spin polarization is slightly larger for the full calculation, but the temporal behavior is very similar to the approximation. The relative difference between the two calculations increases for wider spin initialization. It completely disappears in the case $r_1=0$ ($\alpha=\beta$), even for $r_2>0$, which is consistent with the disappearance of the off-diagonal matrix elements of $\tilde{\boldsymbol D}$ that contain $q_x$.

\section{Comparison with experiment}

To verify the two predicted consequences the finite spatial extension of the initial excitation has, namely, (1) the additional decay rate $D_s q_0^2 w^2 / w'^2$ and (2) the time dependence of $q_0'$ [Eq.~(\ref{eq:q0time})], we compare the model with time- and spatially resolved measurements of the spin dynamics in a (001)-grown GaAs/AlGaAs quantum well. We use the experimental technique described in Ref.~\onlinecite{Walser2012}. In brief, spin polarization is created in the conduction band by a circularly polarized pump laser pulse via the optical orientation effect. A second laser pulse (probe) maps the evolving spin polarization $S_z(x,y,t)$ using the magneto-optical Kerr effect. The pump and probe beams are focused onto the sample surface, where the intensity profile of the probe beam reaches a width of 1\,$\mu$m (Gaussian sigma). The pump beam spot is set to two different sizes: One comparable to the probe beam, and the other twice as large. Spatial maps are recorded by scanning the pump beam over the sample surface. The time evolution is monitored by varying the time delay between pump and probe. The quantum well investigated is characterized by the following parameters: sheet density $n_s=4\cdot10^{15}$\,m$^{-2}$, $\beta_1=4.1\cdot 10^{-13}$\,eVm, $\beta_3=0.7\cdot 10^{-13}$\,eVm and $\alpha=2.8\cdot 10^{-13}$\,eVm.

\begin{figure}[ht]
\includegraphics[width=45mm, angle=270]{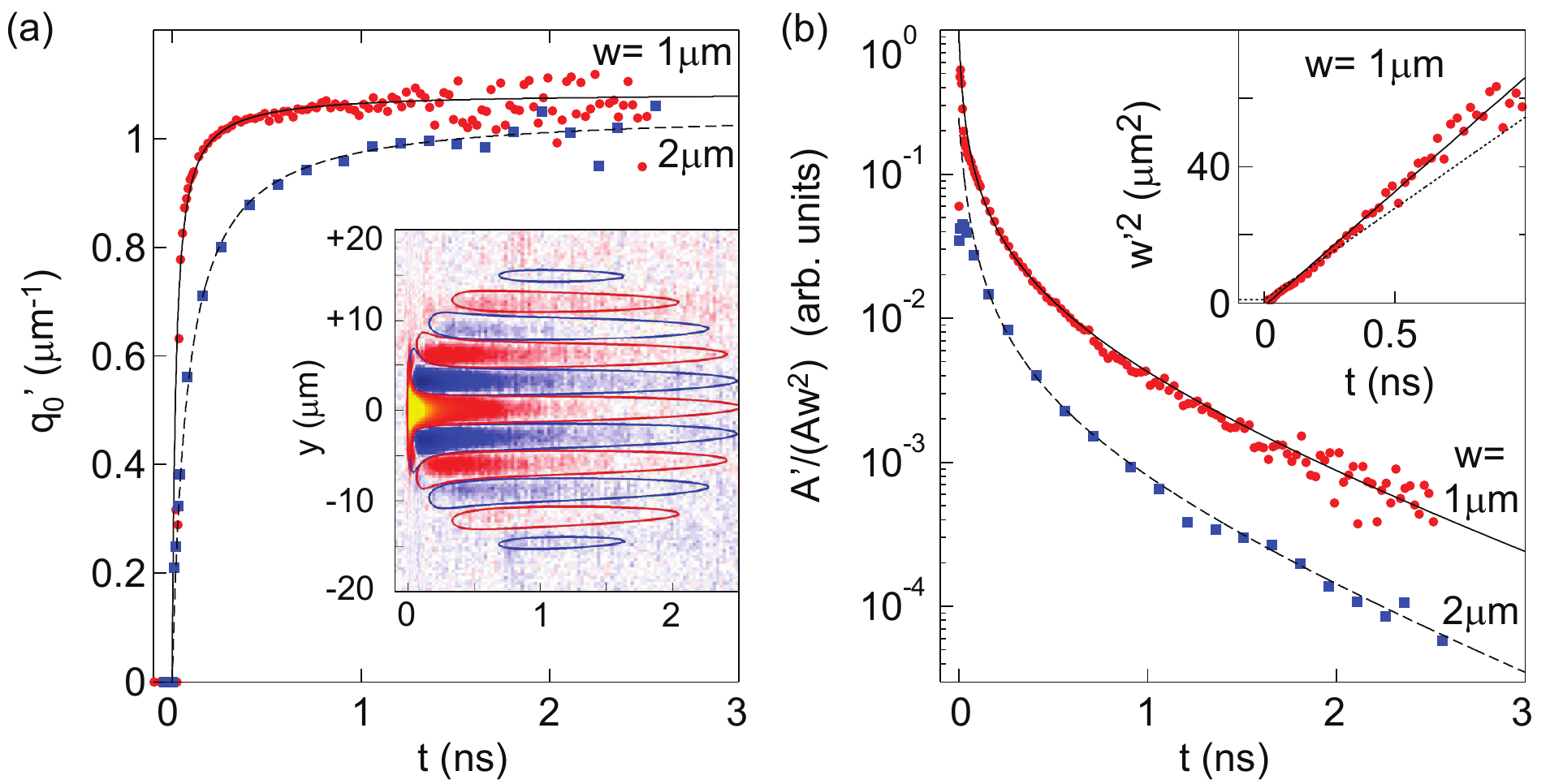}
\caption{\label{fig:fig7} Dynamics of a local spin excitation of size $w=1\,\mu$m (dots) and $2\,\mu$m (squares). The experimentally observed dynamics (symbols) is compared with a theoretical model (curves). In (a) the time dependence of the wave number $q_0'(t)$ and in (b) the spin polarization amplitude $A'/(Aw^2)$ is shown. The color-scale (contour) plot in the inset of (a) shows the measured (modeled) map of $\rho_z(0,y,t)$ for a small pump spot size ($w=1\,\mu$m). The inset in (b) contains the experimentally obtained $w'^2(t)$ (dots) and the fit from the global model (dashed line). The solid line shows a linear fit to $w'^2(t)$ restricted to $t<1$\,ns, yielding a slightly larger diffusion constant of $D_s=334\,$cm$^2$/s, possibly caused by a decrease of $n_s$ with time.}
\end{figure}

The color-scale plot in Fig.~\ref{fig:fig7}(a) shows a map of $\rho_z(0,y,t)$ recorded by scanning the position of the incident pump beam along the $y$-direction at various times $t$ after excitation (for the small pump spot). The contour lines mark the positions of $\rho_z(0,y,t)\approx 0$ of a model based on Eq.~(\ref{eq:2Dmodeapprox}) that is globally fitted to the full data set $\rho_z(0,y,t)$. The parameters $w'$ and $q_0'$ have been modeled according to Eqs.~(\ref{eq:wtime}) and (\ref{eq:q0time}), respectively. In the case of a small (large) excitation spot the following fit parameters were obtained: $w=1.00$ (2.02)\,$\mu$m,  $\Gamma_s^{-1}=1.12$\,ns (1.01),  $q_0=1.08$\,$\mu$m$^{-1}$ (1.05), and $D_s=267$\,cm$^2$/s (262). Note that the values obtained for $w$ are determined by the dynamics of $\rho_z(0,y,t)$ after excitation, specifically by $q_0'(t)$ and $A'(t)$, and were not fixed to a predefined value.

Next we examine the measured dynamics of the wave vector $q_0'$ and the spin polarization amplitude $\rho_z(0,0,t)$ in Figs.~\ref{fig:fig7}(a) and (b).
The symbols show $q_0'$ and $A'$ obtained from individual fits of $A'\exp(-y^2/2w'^2) \times \cos(q_0' y)$ to experimental data at various $t$. The effect of the probe beam size is included in the fit by a convolution of the fit function with a Gaussian of sigma width 1\,$\mu$m. The solid and dashed lines represent the modeled time dependencies. It can be seen from Fig.~\ref{fig:fig7}(a) that $q_0'$ approaches $q_0$ significantly faster in the case of a small spot (dots) than in the case of a large spot (squares), thereby confirming statement (2). Also the measured amplitudes $A'(t)$ are in excellent agreement with the model [Fig.~\ref{fig:fig7}(b)]. For the larger excitation spot, the initial signal decay is stronger than for the smaller spot. This supports the existence of the additional decay as expected from statement (1) above.

\section{Conclusion}

A model has been developed for the spatial and temporal evolution of the spin polarization $\boldsymbol{\rho}(x,y,t)$ in a (001)-oriented quantum well in a zincblende semiconductor. A situation in which Rashba and Dresselhaus SOI are of similar size has been considered. Specifically, in this system we investigated the spin dynamics after an initial local spin excitation with polarization along $z$ and of lateral spatial extension $w$. For a spatially delta-shaped excitation ($w\rightarrow 0)$, the persistent spin helix mode at wave number $q_0$ is fully excited, and $\rho_z$ is given by a helical spin mode with constant wave number $q_0$ and wrapped into a Gaussian envelope that expands diffusively. The spin polarization decays exponentially with a decay rate given by $\Gamma_s=2 D_s {m^*}^2/\hbar^4\left(\left(\alpha-\beta\right)^2+3\beta_3^2\right)$. For finite $w$, there are two important modifications: First, there is an additional decay mechanism that can be described by a time-dependent decay rate $D_s q_0^2 w^2 / w'^2$. This rate suppresses the spin polarization at times $t\gg w^2/2D_s$ by a factor $\exp({-q_0^2 w^2/2})$, which corresponds to the weight of the initial excitation at $q_0$ in Fourier space. Second, the wave number $q_0'$ of the helical spin polarization becomes time dependent and reaches $q_0$ only asymptotically for $t \gg w^2/2D_s$. Both modifications are observed in experimental data obtained by time-resolved Kerr rotation measurements on GaAs/AlGaAs quantum-well samples. These results are also relevant in the case of electrical spin injection from, e.g., ferromagnetic contacts.

An important consequence of our finding is that the determination of $\Gamma_s$ from experimental data requires some care because of the signal suppression and the slower formation of the helical spin mode for finite $w$. As an example, the time scale where $q_0'$ reaches $0.9 q_0$ is $9 w^2/2D_s$, which amounts to 670 \,ps for $w=2$\,$\mu$m in our experiment. This effect is even more pronounced for smaller $D_s$, for example in samples with smaller electron sheet densities. Also the prefactor $w^2/w'^2$ in Eq.~(\ref{eq:2Dmodeapprox}) influences the spin transient significantly.

In the limit of $w \rightarrow \infty$, Eq.~(\ref{eq:2Dmodeapprox}) predicts an exponential spin decay rate of the out-of-plane spin polarization of $D_s q_0^2+\Gamma_s$. This is in agreement with $i\omega(q_x=q_y=0)$ in Eq.~(\ref{eq:dispersion}) and equivalent to $\tfrac{1}{2}(\tau_z^{-1}+\tau_y^{-1})$ with the usual Dyakonov-Perel rates~\cite{Kainz2003}
$\tau_z^{-1}=8D_s{m^*}^2/\hbar^4\left(\alpha^2+\beta^2+\beta_3^2\right)$
and $\tau_y^{-1}=4D_s{m^*}^2/\hbar^4\left((\alpha+\beta)^2+\beta_3^2\right)$. Note that in this limit, the decay rate of a spin polarization along $z$ is not given by $\tau_z^{-1}$ alone, but by the average of $\tau_z^{-1}$ and $\tau_y^{-1}$. This is a consequence of the correlation of position and spin orientation of each individual electron in the balanced SOI situation with $r \ll1$. Even though the spin ensemble decays rapidly because of the spatially broad excitation, individual electron spins form a helix and thereby rotate in the $y$-$z$ plane.

\section*{Acknowledgements}

Financial support from NCCR Nano and NCCR QSIT is acknowledged. We thank R. Allenspach, Y. S. Chen, A. Fuhrer, D. Loss and R. Warburton for fruitful discussions.



\end{document}